\documentclass[twocolumn,showpacs,amssymb,prl,aps,superscriptaddress]{revtex4}
\usepackage{graphicx}
\usepackage{bm}

\begin{document}

\title{Lamellar phase separation and dynamic competition in
La$_{0.23}$Ca$_{0.77}$MnO$_3$}

\author{J. Tao}
\affiliation{Department of Materials Science and Engineering,
University of Illinois at Urbana-Champaign, Urbana, Illinois
61801}

\author{D. Niebieskikwiat}
\affiliation{Department of Physics, University of Illinois at
Urbana-Champaign, Urbana, Illinois 61801}

\author{M.B. Salamon}
\affiliation{Department of Physics, University of Illinois at
Urbana-Champaign, Urbana, Illinois 61801}

\author{J.M. Zuo}
\affiliation{Department of Materials Science and Engineering,
University of Illinois at Urbana-Champaign, Urbana, Illinois
61801}

\date{\today}

\begin{abstract}

We report the coexistence of lamellar charge-ordered (CO) and
charge-disordered (CD) domains, and their dynamical behavior, in
La$_{0.23}$Ca$_{0.77}$MnO$_3$. Using high resolution transmission
electron microscopy (TEM), we show that below $T_{CD}\sim170$ K a
CD-monoclinic phase forms within the established CO-orthorhombic
matrix. The CD phase has a sheet-like morphology, perpendicular
to the ${\bm q}$ vector of the CO superlattice ($a$ axis of the
$Pnma$ structure). For temperatures between $64$ K and $130$ K,
both the TEM and resistivity experiments show a dynamic
competition between the two phases: at constant $T$, the CD phase
slowly advances over the CO one. This slow dynamics appears to be
linked to the magnetic transitions occurring in this compound,
suggesting important magnetoelastic effects.

\end{abstract}


\maketitle

Decades of research have yet to completely understand the full
range of ordered phases that occur in the A$_{1-x}$A'$_x$MnO$_3$
perovskites (A=La, Pr, Nd and A'=Ca, Sr, Ba) \cite{reviews}. For
$x\leq0.5$, a complex competition occurs among ferromagnetic
(FM), paramagnetic (PM), and charge-ordered (CO) antiferromagnetic
(AFM) phases. In this regime, inhomogeneous phase separation (PS)
gives rise to the well-known Colossal Magnetoresistance (CMR)
effect \cite{jin94,dagotto01,fath99}. This regime has been widely
studied and is reasonably well understood in terms of
disorder-induced PS \cite{dagotto01,salamon02}, which produces a
fractal-like morphology of the coexisting domains \cite{fath99}.

Less well examined, and far richer, behavior is found for
$x\geq1/2$. In this region, electron microscopy and other
spectroscopic tools find strong evidence for the appearance of
charge ordering. The CO phases are characterized by the
appearance of superlattice diffraction spots with a wave vector
$q=1-x$, corresponding to real-space ``stripes" with a period
$a/(1-x)$, perpendicular to the $a$ axis of the orthorhombic
perovskite structure ($Pnma$ space group) \cite{cheongJAP}. As
the temperature is lowered below the CO transition at $T_{CO}$,
neutron scattering shows the emergence of AFM ordering of the Mn
moments at $T_N<T_{CO}$ \cite{neutrons66}. Because charge is
discrete, we would expect a robust CO phase to occur at rational
values of $x$; i.e. at $x=1/2$, $2/3$, $3/4$ ... Indeed, for
$x=2/3$ a fully commensurate CO/AFM state appears. Surprisingly,
Pissas and Kallias \cite{pissas03} reported recently that the
$x=3/4$ state is unstable, phase separating into a CO/AFM state
similar to the $x=2/3$ structure, and a C-type AFM phase not
unlike that for $x=0.80$ \cite{neumeier03}. In this paper, we
demonstrate that this PS is dynamic, and results in lamellar
charge-disordered regions that grow in number and size below a
charge disordering temperature $T_{CD}$. Both the structure of
the lamellae, and the appearance of additional Bragg peaks
indicate that this separation also involves a change in crystal
structure. We monitor the evolution of the phases through
time-dependent measurements of resistivity, magnetization, and
high-resolution transmission electron microscopy (TEM) images.

Our polycrystalline sample La$_{0.23}$Ca$_{0.77}$MnO$_3$ was
prepared by the nitrate decomposition route, as described
elsewhere \cite{dario02}, with a final sintering process at
$1500$ $^{\circ}$C for 24h. In-situ temperature dependent dark
field electron imaging and electron diffraction were carried out
using the JOEL 2010F ($200$ kV) transmission electron microscope
under the experimental conditions previously described
\cite{jing04}. Magnetization ($M$) measurements as a function of
temperature were performed using a commercial SQUID magnetometer.
The $M$ data were obtained by warming the sample in a magnetic
field $H=100$ Oe after zero-field-cooling (ZFC) and field-cooling
(FC) processes from $300$ K to $10$ K. The resistivity ($\rho$)
was measured by the usual four probe method.

Figure 1 shows the temperature driven evolution of the
microstructure of a selected region of the sample, recorded by
high resolution dark field imaging using a fundamental reflection
and the two corresponding CO superlattice spots. At $T=175$ K
[Fig. 1(a)] the stripe-like contrast formed by the interference
between these three reflections is evident
\cite{cheongJAP,jing04}. In magnetization measurements, the onset
of the CO phase is indicated by a peak in the $M(T)$ curve, as
shown in the inset of Fig. 2. The occurrence of this peak was
ascribed to the freezing of the double exchange interaction due
to the localization of the charge carriers below $T_{CO}$
($\sim220$ K in our case). However, contrary to the nearly
uniform CO phase present at $175$ K, in Fig. 1(b) ($T=145$ K) it
can be seen that some non-striped ``bands" (sheets in 3D)
nucleate inside the CO regions. The disappearance of the CO
stripes indicates that these bands are undergoing a transition to
a charge-disordered (CD) phase, which starts to occur at
$T_{CD}\sim170$ K. As the temperature is lowered the CD regions
grow in both number and size, and the image in Fig. 1(c) at
$T=99$ K (the lowest achievable temperature for our TEM) exhibits
a substantial amount of the CD phase. The picture presented in
Fig. 1(d) shows a larger area ($\sim0.6$ $\mu$m wide) indicating
that the appearance of this CD phase is not a local effect, but
is spread throughout the sample with domain sizes ranging from
$\sim10$ to $30$ nm wide.

Several important observations can be made from the images shown
in Fig. 1. First of all, the CD phase cannot be attributed to a
chemical segregation or composition variations in the sample. The
CD domains occupy the same regions where the homogeneous CO phase
was present at higher temperatures. Moreover, repetitions of the
same experiment show similar images, but the CD areas do not
always appear at the same locations in the sample, i.e. the CD
phase is not pinned to defects.

Secondly, the morphology of the lamellar phase coexistence is
striking. The minority CD domains appear as sheets embedded within
the CO phase, where the domain walls separating both phases are
nearly perpendicular to the $a$ axis of the orthorhombic
perovskite structure. This consideration indicates that, contrary
to previous results in other phase separated manganites
\cite{dagotto01,salamon02}, random disorder hardly could be
responsible for the coexistence of phases observed in
La$_{0.23}$Ca$_{0.77}$MnO$_3$. In this kind of phase separation,
the elastic energy must play a fundamental role in order to
produce such smooth domain walls with their normals determined by
the [100] propagation direction of the CO structural distortion.

In order to determine the nature of the CD phase we studied the
electron diffraction (ED) pattern of the sample at $T=175$ K
[inset of Fig. 1(a)] and $T=99$ K [inset of Fig. 1(c)]. It can be
observed that around the fundamental reflections [(104) is shown]
the superlattice spots of the CO phase show up at both
temperatures. These CO spots appear right below $T_{CO}$, with a
wave vector $q\approx0.23$(1) indicative of the doping level
$x\approx0.77$. In addition to these spots, the white arrows in
the inset of Fig. 1(c) signal the presence of another set of
reflections near the strong fundamental one. These reflections
are totally absent for $T>T_{CD}$, indicating that they are
related to the CD phase. These satellite reflections were indexed
to a monoclinic phase sharing the common $a$ and $b$ axis with
the CO-orthorhombic phase, but with the $c$ axis tilted by an
angle of $\sim1.3^{\circ}$ from the orthorhombic structure. The
two spots at each side of the fundamental reflection correspond
to different monoclinic domains with the positive and negative
tilting angle, respectively.

Therefore, below $T_{CD}$ we observe the coexistence of the
CO-orthorhombic (CO-O) and CD-monoclinic (CD-M) phases. The
observation of the secondary CD-M phase is consistent with neutron
scattering measurements \cite{pissas03}, where $\sim25\%$ of the
volume of La$_{1/4}$Ca$_{3/4}$MnO$_3$ samples was found to belong
to the $P2_1/m$ monoclinic space group at low $T$. Also the
temperature $T_{CD}\approx170$ K coincides with the appearance of
extra magnetic reflections in the neutron diffraction pattern.
These reflections were associated to the monoclinic phase which
adopts the C-type AFM structure \cite{neumeier03}.

In our magnetization curves (inset of Fig. 2) the proximity of the
prominent ``CO peak" hinders the observation of the AFM transition
related to the appearance of the CD-M phase. However, the $M(T)$
data present some other curious features. As the temperature
decreases, the splitting between the ZFC and FC curves becomes
significant below $T_S\approx130$ K, and at $T_C\approx64$ K a
small but clear magnetization increase occurs. While $T_S$
indicates the PM to AFM transition of the CO phase, the weak FM
moment showing up at $T_C$ is related to the monoclinic phase
\cite{dario04}. Figure 2 plots the curve of the reduced magnetic
moment $m(T)=M_{FC}/M_{ZFC}$, where $M_{FC}$ and $M_{ZFC}$ are the
magnetization values for the FC and ZFC processes, respectively.
At high temperature the reduced moment remains close to $m=1$, as
expected for a normal PM phase. On the other hand, as soon as the
CO phase is formed $m$ starts to increase. Since immediately below
$T_{CO}$ the CO phase does not adopt any magnetic order, the value
$m>1$ indicates the presence of defects within the CO background
with a ``glassy" response to the application of a magnetic field.
The other two magnetic transition temperatures, $T_S$ and $T_C$,
are characterized by abrupt breaks in $m(T)$.

The most curious characteristic of this CO-CD phase coexistence
precisely occurs between the two magnetic transitions, i.e for
$T_C<T<T_S$. In this temperature range, we observed an anomalous
dynamic regime by measuring the temporal relaxation of the
resistivity. Figure 3 presents $\rho$ data as a function of time
for selected temperatures, recorded over periods of $\sim3$ hours.
For each curve, the sample was warmed up to $300$ K and then
cooled back at $4$ K/min and stabilized at the measuring
temperature within $\pm2$ mK. For $T_S<T<T_{CO}$ the resistivity
increases with time. This effect is present even above $T_{CD}$,
indicating that it is not related to the phase coexistence but is
a property of the CO-O phase itself. Moreover, we observed the
same effect in the CO-O phase of La$_{1/3}$Ca$_{2/3}$MnO$_3$
samples, where the CD-M phase was not observed in the TEM studies
\cite{jing04}. For $x=0.77$, the increase of resistivity with time
is correlated with an increase in the intensity of the CO
superlattice reflections (not shown), which indicates that the CO
phase is becoming more ordered with time \cite{dario04}. This
behavior supports the existence of mobile microscopic defects in
the CO-O phase, which are responsible for the aforementioned
glassy response observed in $m(T)$ for $T<T_{CO}$. The ability of
these defects to migrate outside the CO volume provides the
mechanism for the improvement of ordering with time.

In the inset of Fig. 3 we show the logarithmic relaxation rate
$S=d(\log \rho)/d(\log t)$ as a function of temperature, obtained
from the time interval between $10$ and $20$ minutes. The positive
values of $S$ for $T>T_S$ relate to the increase of $\rho$ with
time. A similar time evolution ($S>0$) is observed for $T$ below
$T_C=64$ K. Surprisingly, in the intermediate region $T_C<T<T_S$
the relaxation rate changes sign ($S<0$) and the resistivity
decreases with time. Remarkably, the changes in the dynamical
behavior coincide with the magnetic transitions at $T_S$ and
$T_C$. In order to gain some insight on the dynamic properties of
this compound we obtained dark field images of a given region of
the sample at different times. As shown in Fig. 4, when the sample
is cooled to $T=99$ K (at approximately $4$ K/min) the phase
separation pattern evolves with time. Figure 4(a) shows that
immediately after the sample is stabilized at $99$ K the CD-M
phase occupies a small part of the volume. However, some regions
of the sample undergo a transition from the CO-O to the CD-M
phase (see the framed region for example), and the volume fraction
of the latter keeps increasing with time. At 30 minutes after the
temperature is reached [Fig. 4(d)] the CD-M phase occupies an
appreciable fraction of the area of the image. Beyond $t=30$
minutes the relaxation becomes very slow, and no obvious change
was observed in the images. It is worth mentioning that similar
relaxation effects were also observed at $T=115$ K. At $T=125$ K
the time evolution of the pattern is very slow and at $T=135$ K
($>T_S$) it becomes unobservable. Thus, all these structural
changes have an absolute correlation with the resistivity
relaxation.

The above results present a consistent scenario of the complex
dynamic competition between the CO and CD phases for $T$ below
$T_S$. It is well known that a charge-ordered phase provides a
strong insulating state due to the localized nature of the charge
carriers on well-defined Mn sites. Differently, the particular
C-type AFM structure of the CD-M phase consists of
antiferromagnetically coupled FM chains, inside which the double
exchange interaction \cite{zener51} favors the delocalization of
the charge carriers \cite{neumeier03}. As a result, even though
the AFM nature of the CD-M phase also produces an overall
insulating state, the carriers' motion along these
one-dimensional paths leads to a reduction of the resistivity.
Therefore, contrary to the case of the upward relaxation, for
$T<T_S$ the negative $d\rho/dt$ must be due to the slow growth of
the CD-M phase in the CO-O matrix. For $T<64$ K, the return of
the relaxation rate towards positive values indicates that the
motion of the boundaries between the two phases is stopped again.

A peculiar feature about the interplay between magnetic,
electronic, and structural degrees of freedom in
La$_{0.23}$Ca$_{0.77}$MnO$_3$ is that the occurrence of a magnetic
phase transformation is able to induce a change in the dynamic
properties of the phase coexistence. The new magnetic order that
appears below $T_S\approx130$ K triggers the motion of the domain
walls that separate the CO-O and CD-M domains. At lower
temperature, the appearance of a FM component is able to freeze
this motion again, driving the system back to a static phase
separated state. This behavior implies the existence of strong
magnetoelastic effects \cite{natureelasticity}. Indeed, a
magnetostrictive response was already observed \cite{neumeier03}
at the N\'eel temperature of the (orthorhombic) G-type AFM
structure in La$_{0.12}$Ca$_{0.88}$MnO$_3$. This magnetic
transition was also found to have an appreciable impact on the
structural properties of the also present C-type AFM phase. It was
previously argued that the large strain produced by the CO state
on the borders of the domains greatly affects the physical
properties of phase separated manganites, producing
martensitic-like features \cite{martensitic}. In the case of our
La$_{0.23}$Ca$_{0.77}$MnO$_3$ sample, a change in the magnetic
structure of the coexisting phases would produce a change in the
strain fields at the boundaries between them, thus affecting the
delicate energy balance of the system and allowing (or precluding)
the migration of the domain walls.

Summarizing, the close correlation between the high resolution
TEM images and the physical properties allowed us to obtain a
complete picture about the phase separated state in the
La$_{0.23}$Ca$_{0.77}$MnO$_3$ manganite. Below the CO transition
at $T_{CO}\sim220$ K the high resolution images show the typical
CO stripes fully formed in the whole sample. However, on further
cooling below $T_{CD}\sim170$ K some non-CO laminated regions
with a monoclinic structure form within the CO-orthorhombic
matrix. The particular morphology of the charge-disordered domains
and the simultaneity of magnetic and dynamic transitions observed
at lower temperatures indicate that the CO-CD phase coexistence
in this manganite must be driven by magnetoelastic effects rather
than being dictated by disorder. The fundamental question that
remains open is what is the real ground state of the electron
doped manganites. For $x\approx 0.77$, contrary to the widely
expected CO phase, the present results support a monoclinic
C-type-AFM ground state. The realization of such phase seems to
be only impeded by the inability of the domain walls to move out
of the sample fast enough, stopped by the strain fields induced
by the high-temperature CO phase.

This material is based upon work supported by the U.S. Department
of Energy, Division of Materials Sciences under award No.
DEFG02-91ER45439, through the Frederick Seitz Materials Research
Laboratory and the Center for Microanalysis of Materials at the
University of Illinois at Urbana-Champaign. J.T. was supported by
DOE under award No. DEFG02-01ER45923.

\vspace{1cm}

\noindent FIG 1. (a) to (c) Temperature evolution of the dark
field images of the same region of the sample. A large area view
of the CO-CD phase separation at $T=99$ K is shown in (d). The
inset in (a) shows the two CO satellite reflections around the
(104) fundamental peak ($T=175$ K). The two additional reflections
observed at $T=99$ K [white arrows in the inset of part (c)]
correspond to the CD phase and were indexed to a monoclinic
structure.

\vspace{1cm}

\noindent FIG 2. Inset: Magnetization as a function of
temperature measured with a magnetic field $H=100$ Oe after
zero-field-cooling (ZFC) and field-cooling (FC) processes. Main
panel: $m=M_{FC}/M_{ZFC}$ is the ratio between the magnetization
obtained after FC and ZFC, respectively.

\vspace{1cm}

\noindent FIG 3. (Color online) Time dependence of the electrical
resistivity of La$_{0.23}$Ca$_{0.77}$MnO$_3$ at different
temperatures. Inset: logarithmic relaxation rate $S=d(\log
\rho)/d(\log t)$ as a function of temperature, taken from the
time interval $10$ min$<t<20$ min.

\vspace{1cm}

\noindent FIG 4. Temporal evolution of the dark field images of
the sample. (a) Initial stage; (b), (c), and (d) were taken after
1, 5, and 30 minutes, respectively. The frames show a region
where the CD phase is expanding with time.

\end{document}